\documentclass[%
 reprint,
 amsmath,amssymb,
 aps,
]{revtex4-2}

\usepackage{graphicx}
\usepackage{dcolumn}
\usepackage{bm}
\usepackage{hyperref}
\usepackage{amsthm}
\usepackage{slashed}
\usepackage{amsmath}
\usepackage{amssymb}
\usepackage{cancel}
 \usepackage[normalem]{ulem}
\usepackage{amsfonts}
\usepackage{mathrsfs}
\usepackage{mathtools}
\usepackage{xcolor}

\def\diff{\textrm{d}}

\def\dj{d\kern-0.4em\char"16\kern-0.1em}
\def\Dj{\mbox{\raise0.3ex\hbox{-}\kern-0.4em D}}

\begin{document}

\title{Holographic entanglement entropy in Chern-Simons gravity with torsion}

\author{Du\v{s}an \Dj or\dj evi\'{c}}
\author{Dragoljub Go\v{c}anin}
\email{dragoljub.gocanin@ff.bg.ac.rs}
\affiliation{%
Faculty of Physics, University of Belgrade\\ Studentski Trg 12-16, 11000 Belgrade, Serbia
}%

\begin{abstract}
Holographic entanglement entropy is a key concept linking quantum information theory and gravity. Since the original conjecture of Ryu and Takayanagi, holographic entanglement entropy has been generalized beyond Einstein--Hilbert gravity to include higher-curvature corrections. In most existing generalizations, however, it is implicitly assumed that the bulk spacetime geometry is Riemannian, \emph{i.e.}\ torsion-free. Here we propose a prescription for incorporating torsion into holographic entanglement entropy in the boundary theory dual to five-dimensional Chern--Simons gravity. We argue that the entanglement entropy acquires an additional universal divergent term proportional to the logarithm of the UV cutoff, and that this term is generated solely by torsion.
\end{abstract}

\maketitle

\section{Introduction}

Entanglement entropy is one of the key concepts linking quantum information theory, condensed matter physics, and quantum gravity \cite{Chen:2021lnq}. In quantum field theory (QFT), the entanglement entropy is notoriously difficult to compute in most cases. Due to short-range correlations in the vicinity of the entangling surface $\Sigma$ that separates two spatial regions (subsystems) on a given Cauchy slice, the entanglement entropy of a $D$-dimensional QFT has ultra-violet (UV) divergences, 
and therefore must be regulated. While the leading power-law divergences are highly sensitive to the choice of UV regulator $\epsilon$, subleading terms can exhibit a universal character and thereby encode the intrinsic information about the underlying QFT state, a canonical example being an even-dimensional conformal field theory (CFT), for which the entanglement entropy contains a universal logarithmic divergence \cite{Lewkowycz:2012qr}. 

Ryu and Takayanagi (RT) proposed a radical new way of thinking about entanglement entropy in holographic QFT~\cite{Ryu:2006ef, Rangamani:2016dms}. They argued that the entanglement entropy of a spatial region $\mathcal{A}$ in a $D$-dimensional QFT admits a holographic interpretation as the area (in appropriate units) of a codimension-$2$ extremal surface $\gamma_{\mathcal A}$ embedded in a $(D+1)$-dimensional bulk spacetime and anchored on the entangling surface $\Sigma=\partial\mathcal A$. The RT holographic entanglement entropy (HEE) formula can be viewed as a generalization and reinterpretation of the Bekenstein--Hawking black-hole entropy formula, and it has become a central tool in AdS/CFT \cite{Maldacena:1997re} and holography-based QFT research.
The original RT proposal was somewhat restricted, and several refinements soon followed \cite{Dong:2013qoa, deBoer:2011wk, Erdmenger:2014tba, Anastasiou:2021jcv}. Yet, in all of them---as is generally the case in holographic considerations---the bulk theory is taken to be classical gravity based on (pseudo)-Riemannian geometry, where the affine structure is fixed to be the Levi--Civita connection and torsion is set to zero. While this is the standard choice in Einstein gravity, it is not required \emph{a priori}: there exist well-motivated extensions of general relativity in which torsion plays a distinctive dynamical role \cite{Blagojevic:2002du}.
Such theories are naturally formulated in Riemann--Cartan (RC) geometry, where metric and affine (parallelism) structures are treated as independent. In the so-called first-order formalism, one works with two independent dynamical 1-form fields: the vielbein (frame field) $e^{a}$ and the spin connection $\omega^{ab}$ -- the gauge field for local $SO(D\!-\!1,1)$ Lorentz transformations. The curvature and torsion 2-forms are defined as
\begin{equation}
R^{ab} = \mathrm{d}\omega^{ab} + \omega^{a}{}_{c}\wedge \omega^{cb}\,,\qquad
T^{a} = \mathrm{d}e^{a} + \omega^{a}{}_{b}\wedge e^{b}\,.
\end{equation}
Although RC holography remains comparatively less explored, there are indications that extending holographic methods to torsion-full bulk geometries may be fruitful \cite{Banados:2006fe, Gallegos:2020otk}. For example, in condensed-matter settings, torsion can provide an effective description of spin-transport phenomena \cite{Gallegos:2020otk}, and even play a role in explaining certain aspects of holographic conductivity \cite{Dordevic:2025akw}. A holographic account of such systems motivates understanding how torsion modifies entanglement entropy from the bulk perspective.
In this paper, we consider, in a holographic setting, a particular solution of 5-dimensional CS gravity in the bulk, namely the AdS$_5$ with axial torsion, and propose a prescription for introducing torsion in the computation of the entanglement entropy of its 4-dimensional holographic dual. Our conjecture is guided by previous results on 5-dimensional Riemannian (torsion-free) Lovelock theory~\cite{deBoer:2011wk} and by Wald's formula for the black-hole entropy in Chern--Simons (CS) gravity with torsion \cite{Gallegos:2020otk}. 
We show that a  universal torsion-induced logarithmically divergent term of the entanglement entropy of a 4-dimensional CFT can be reproduced holographically in a 5-dimensional CS gravity setting. 

\section{Torsion-induced entanglement entropy}

To the best of our knowledge, a study of entanglement entropy in a QFT defined on a background spacetime with non-vanishing torsion has not yet been carried out. However, there is some related work on HEE based on the first-order formulation of gravity that we should comment on. 
In general,  
RC geometry provides a natural setting for formulating gravity as a gauge theory. A well-known example is Einstein-Hilbert (EH) gravity (with negative cosmological constant) in three-dimensions, which can be formulated as the CS theory for the $SO(2,2)$ gauge group,
\begin{equation}
S_{\text{CS}}[A]=\frac{k}{4\pi}
\int
\operatorname{Tr}
\left(
A \wedge \textrm{d}A
+
\frac{2}{3}
A \wedge A \wedge A
\right)\,,
\end{equation}
where $A \in \mathfrak{so}(2,2)$ is the gauge connection and $k$ is the CS level.
The gauge connection can be decomposed into vielbein and spin connection as
\begin{equation}
A_\mu = \frac{1}{\ell} e^{a}_\mu\, P_a + \frac{1}{2}\omega^{ab}_{\mu} M_{ab}\,,
\end{equation}
where $M_{ab}$ are the Lorentz generators and $P_a$ are the translation generators
of $\mathfrak{so}(2,2)$. The length parameter $\ell$ is the AdS radius. It is often convenient to write $SO(2,2)\approx SL(2,\mathbb{R})\times SL(2,\mathbb{R})$, in which case the connection $A$ can be decomposed into two independent parts,
\begin{equation}
A = \left(\omega^{a} +  \frac{1}{\ell}e^{a}\right) J^{+}_{a}\,,
\qquad
\bar{A} = \left(\omega^{a} -\frac{1}{\ell} e^{a}\right) J^{-}_{a}\,,
\end{equation}
with $\omega_{\mu}{}^{a}\,\epsilon_{ab}{}^{c}
=
\omega_{\mu}{}^{c}{}_{b}$ and the $SL(2,\mathbb{R})$ generators $J^{\pm}_{a}
=
\tfrac{1}{2}
\left(
M_{a} \pm \, \ell P_{a}
\right)$, where $M_{a}\omega^{a}=\tfrac{1}{2}\omega^{ab} M_{ab}$. It is customary to put $\ell=1$. The EH action can now be written as
\begin{equation}\label{action3D}
S_{\text{EH}}
=
S_{\text{CS}}[A]
-
S_{\text{CS}}[\bar{A}] \,.
\end{equation}
It was shown in \cite{Ammon:2013hba} that in the first-order formulation of gravity, the role of the RT surface can be played by a gravitational Wilson line. 
This result is then used to generalize the computation of the HEE to higher-spin gravity theories. 
In particular, the path that the Wilson line takes in the bulk is given by 
\begin{equation}\label{WL}
    \frac{\diff}{\diff s}\left( (A - \bar{A})_\mu \frac{\diff x^\mu}{\diff s} \right)
+ [\bar{A}_\mu, A_\nu]\,
\frac{\diff x^\mu}{\diff s}\frac{\diff x^\nu}{\diff s}
= 0\,,
\end{equation}
which is precisely the geodesic equation for the curve $x^{\mu}(s)$ in a spacetime described by $e^{a}_{\mu}$ and $\omega^{ab}_{\mu}$, i.e. 
\begin{equation}\label{geod}
\frac{\diff }{\diff s}
\left(
e^{a}_{\mu}\,
\frac{\diff x^{\mu}}{\diff s}
\right)
+
\omega^{a}_{\mu b}\,
e^{b}_{\nu}\,
\frac{\diff x^{\mu}}{\diff s}
\frac{\diff x^{\nu}}{\diff s}
=
0\,.
\end{equation}
We point out that, although the first-order formulation is assumed in the derivation of this result, the spin connection appearing in (\ref{geod}) is purely Riemannian, because in 3-dimensional CS gravity torsion vanishes on-shell. One can nevertheless consider more general models, again based on the CS description, that have a proper RC bulk as a solution. For example, this is the Mielke-Baekler (MB) model of 3-dimensional gravity \cite{Mielke:1991nn}, with two equal boundary central charges, and action that consists of the standard 3-dimensional CS action plus the so-called translational CS term that explicitly involves torsion \cite{Blagojevic:2004fu},
\begin{equation}\label{MB}
k\int\varepsilon_{abc}\left(R^{ab}\wedge e^c +\frac{1}{3}e^a\wedge e^b\wedge e^c\right)+\alpha k\int T^a\wedge e_a\,.
\end{equation}
MB theory can be formulated as CS theory of gravity \cite{Blagojevic:2003vn}. In the case of \eqref{MB}, the total action takes a form similar to \eqref{action3D}, but with the gauge connection 
\begin{equation}\label{cudnakoneksijadifferent}
    A=\left(\omega^a+q\, e^a\right)J_a^{+}\,,\hspace{5mm} \bar{A}=\left(\omega^a+\bar{q}\,e^a\right)J_a^{-}\,,
\end{equation}
where $q=\frac{\alpha}{2}+\frac{1}{l}$ and $\bar{q}=\frac{\alpha}{2}-\frac{1}{l}$, with $l^{-2}=1-\frac{3\alpha^{2}}{4}$ being the effective Riemannian AdS radius.
Interestingly, equation (\ref{WL}) reduces to the geodesic (not autoparallel) equation (\ref{geod}) Namely, the solution of the MB theory is not the standard AdS$_3$ because, although the metric remains AdS$_3$, the affine structure changes, and we have an RC spin connection and a non-vanishing torsion. he spin-connection takes the form $\omega^{a}=\Tilde{\omega}^a(e)+\frac{\alpha}{2}e^a$, and we see that \eqref{cudnakoneksijadifferent} takes the standard form once expressed in terms of Riemannian, torsion-free part of the spin connection, which in turn appears in (\ref{geod}). 

Hence, at least in the MB theory, the generalization of the RT proposal seems to be straightforward (assuming that the entanglement entropy is still given by the gravitational Wilson loop with the same boundary conditions as in \cite{Ammon:2013hba}); we simply forget about the affine structure and compute the HEE as in the Riemannian case. This is actually consistent with some other results. Namely, the boundary theory is described by two equal central charges $c_R=c_L=24\pi kl$ and it is defined on a Riemannian manifold, so it is expected that the entanglement entropy should be given by the standard CFT$_2$ result: 
\begin{equation}\label{claim1}
    S_{\text{EE}}=\frac{3}{2G_N\left(1-\frac{3\alpha^2}{4}  \right)^{1/2}} \ln \left(\frac{L}{\varepsilon}\right),
\end{equation}
with $k=\frac{1}{16\pi G_N}$. Furthermore, the entropy of the BTZ black hole in this model is not modified due to torsion, with parameter $\alpha$ entering the entropy formula only through the effective AdS radius $l$ \cite{Blagojevic:2006jk, Klemm:2007yu}. Finally, this type of modification is also present in the AdS/BCFT analysis from \cite{Dordevic:2024ziw}.

Now we move to the more challenging and more important example of 5-dimensional CS gravity \cite{hassaine2016chern}
\begin{align}
   &S_{\textrm{CS}}^{(5)}= k\int\varepsilon_{abcde}\bigg(\frac{1}{\ell}R^{ab}\wedge R^{cd}\wedge e^{e}\\
   &+\frac{2}{3\ell^{3}}R^{ab}\wedge e^{c}\wedge e^{d}\wedge e^{e}+\frac{1}{5\ell^{5}}e^{a}\wedge e^{b}\wedge e^{c}\wedge e^{d}\wedge e^{e}\bigg)\,.\nonumber
\end{align}
Unlike for the 3-dimensional MB model (\ref{MB}), bulk torsion of the 5-dimensional CS gravity induces a non-trivial boundary torsion in the holographic dual through the Fefferman-Graham (FG) expansion. 
Consider, for example, a 5-dimensional CS gravity black hole solution from \cite{Canfora:2007xs, Giribet:2014hpa, Andrianopoli:2021qli}, 
\begin{equation}\label{metrika}
    \diff s^2=-\left( \frac{r^2}{\ell^2}-\mu \right)\diff t^2+\frac{\diff r^2}{\left( \frac{r^2}{\ell^2}-\mu \right)}+r^2\diff \Sigma_3^2\,,
\end{equation}
with non-vanishing axial torsion  
\begin{align}\label{torzija}
T^{0}=T^{1}=0,\hspace{0.5cm}
    T^i=-\frac{\mathcal{C}}{r}\varepsilon_{ijk}e^j\wedge e^k\,,
\end{align}
where the free parameter $\mathcal{C}$ measures the strength of torsion. There are no restrictions for the three-manifold $\Sigma_{3}$, apart from the fact that it is fixed for all values of the radial coordinate. We will not distinguish lower and upper Lorentz indices $i,j,k=2,3,4$. We set $\ell=1$.
In the following we will consider the $\mu=0$ case of (\ref{metrika}) with flat $\Sigma_{3}=\mathbb{R}^{3}$, together with (\ref{torzija}), which represents the AdS$_5$ bulk with axial torsion that induces flat boundary metric and non-vanishing boundary torsion. 
 
In terms of the FG radial coordinate $\rho$ defined by $r= \left(  \frac{1}{\sqrt{\rho}}+\frac{\mu}{4}\sqrt{\rho}\right)=\frac{1}{\sqrt{\rho}}$, the generic FG expansion for the bulk vielbein and spin-connection are given by \cite{Banados:2006fe, Cvetkovic:2017fxa}
\begin{align}\label{FGexp}\nonumber
e^1&=-\frac{\diff \rho}{2\rho}\,, \hspace{0.5cm}e^a=\frac{1}{\sqrt{\rho}}(\overline{e}^a+\rho \overline{k}^a)\,,\\
\omega^{a1}&=\frac{1}{\sqrt{\rho}}(\overline{e}^a-\rho \overline{k}^a)\,,\hspace{0.4cm}
\omega^{ab}=\overline{\omega}^{ab}\,,
\end{align}
where the boundary ($\rho=0$) fields are denoted by a bar sign, and ``non-radial" indices $a$, $b$ take values from $\{0,2,3,4\}$; index 1 is the radial index. 
Having the FG expansion of the bulk fields, we can derive 
the boundary torsion (in the boundary Cartesian coordinates), 
\begin{align}\label{torsion_bound}
\overline{T}^0=0, \hspace{0.5cm }
\overline{T}^i=\mathcal{C} \varepsilon_{ijk}\diff x^j\wedge \diff x^k\,.
\end{align}
and the boundary RC curvature, 
\begin{align}
\bar{R}^{0i}=0, \hspace{0.5cm}
\bar{R}^{ij}=-\mathcal{C}^{2}\;\diff x^{i}\wedge\diff x^{j}\,.
\end{align}
The RC Ricci scalar is then $\bar{\mathcal{R}}_{\text{RC}}=-6\mathcal{C}^{2}$ and it exists solely due to torsion. 

The boundary dual of 5-dimensional CS gravity is a 4-dimensional CFT and, therefore, has anomalous trace of the holographic stress-energy tensor. 
While for a generic four-dimensional CFT the trace contains two types of contribution,
the CS gravity dual has only one anomalous contribution of the form \cite{Banados:2004zt}
\begin{equation}\label{anomm}
\overline{e}^a\tau_a=k\varepsilon_{abcd}\bar{R}^{ab}\bar{R}^{cd}\,.
\end{equation}
This formula also holds for CS gravity with torsion \cite{Banados:2006fe}; one simply has to regard $\bar{R}^{ab}$ as RC curvature instead of the usual Riemann curvature. 
CS gravity is usually not considered as a candidate for bulk theory, as its boundary dual does not have the $c$-type central charge, and it is not a unitary QFT. However, our main interest in CS gravity comes from the fact that it is one of the rare examples where we can understand the role of torsion in a holographic setting. 
Let us first discuss the torsion-free case. As explained in \cite{Lewkowycz:2012qr},
in 4-dimensional CFT with $c=0$ (the vanishing central charge of the trace anomaly), there is a universal divergent contribution to the entanglement entropy of the form 
\begin{equation}\label{See4d}
    S_{\text{EE}}[\mathcal{A}] = \frac{\ln(\epsilon / L)}{2\pi}
\int_{\Sigma} d^{2}\sigma \sqrt{h}\,\;a\,\mathcal{R}_{\Sigma}\,,
\end{equation}
where $h_{\mu\nu}$ is the induced metric and $\mathcal{R}_{\Sigma}$ is the induced Ricci scalar on the entangling surface $\Sigma$. There is a UV cut-off $\epsilon$ and a non-vanishing central charge $a$; $L$ is some characteristic length scale of the system.   
Given that in the boundary dual to CS gravity with torsion, the trace anomaly is computed by a simple substitution of the Riemann curvature with the Riemann-Cartan one, it seems natural to expect that the same formula for the entanglement entropy (\ref{See4d}) also holds when torsion is involved if we simply substitute $\mathcal{R}_{\Sigma}\to \mathcal{R}^{(\text{RC})}_{\Sigma}$, \textit{i.e.}, by the RC Ricci scalar (induced on $\Sigma$). 

For that matter, let us perform a concrete computation by taking a 2-sphere with radius $\mathfrak{R}$ as the entangling surface $\Sigma$. 
The induced Ricci scalar on a hypersurface with a unit normal $n^\mu$ can be computed using the Gauss-Codazzi formula,
\begin{equation}\label{GC}
\mathcal{R}_{\Sigma}=h^{\mu\rho}h^{\nu\sigma}R_{\mu\nu\rho\sigma} +K^2-K_{\mu\nu}K^{\mu\nu}\,,
\end{equation}
where $h^{\mu\nu}=g^{\mu\nu}-n^\mu n^\nu$ and $K_{\mu\nu}=h^{\rho}_\mu h^{\sigma}_\nu\nabla_{\rho} n_{\sigma}$\,. 
An immediate question is how to define the induced RC Ricci scalar. We can use the results of \cite{Erdmenger:2022nhz}, where it was shown that in the RC case we can simply generalize the Riemannian Gauss-Codazzi equation by taking the RC curvature tensor instead of the Riemannian one, and including the contorsion term in the covariant derivative $\nabla_{\rho}n_{\sigma}$. 
Starting from the line element in spherical coordinates
\begin{equation}
    \diff s^2=\diff r^2+r^2\left(\diff\theta^2+\sin^2\theta\diff\varphi^2  \right)\,,
\end{equation}
we have $n_\mu=(1,0,0)$. For our AdS$_5$ bulk with axial torsion, the holographic dual is defined on a 4-dimensional flat background with torsion (\ref{torsion_bound}) induced from the bulk. The contorsion tensor at the boundary is given by $\kappa_{\alpha\beta\gamma}=\mathcal{C}r^2\sin^2\theta\,{\varepsilon}_{\alpha\beta\gamma}$, where $\{\alpha,\beta,\gamma\}$ are spatial indices. The induced RC Ricci scalar is then
\begin{equation}
\mathcal{R}^{(\text{RC})}_{\Sigma}=\frac{2}{\mathfrak{R}}-4\mathcal{C}^2\,.    
\end{equation}
Plugging back into (\ref{See4d}) this yields a universal torsion-induced divergent contribution of the form
\begin{equation}\label{ln1}
 \mathfrak{R}^2\mathcal{C}^2\ln \epsilon\,.
\end{equation}
This is the result we expect to obtain starting from the HEE of 5-dimensional CS gravity properly generalized to include torsion.   
For 5-dimensional CS gravity under the torsion-free constraint the HEE is given by (modulo boundary terms)
\begin{equation}\label{EE}
S_{\text{HEE}}[\mathcal{A}]=2\pi k \int_{\gamma^{0}_{\text{ext}}}d^{3}y\; \sqrt{h}\left(1+\frac{1}{2}\mathcal{R}_{\gamma^{0}_{\text{ext} }}\right)\,,
\end{equation}
where $\gamma^{0}_{\text{ext}}$ is the RT codimension-2 extremal surface anchored at region $\mathcal{A}$ and  $\mathcal{R}_{\gamma^{0}_{\text{ext}}}$ is the torsion-free Ricci scalar induced on $\gamma^{0}_{\text{ext}}$. It is also known that there is a close relation between HEE (in the sense of a generalized RT proposal) and Wald's black hole entropy, which will serve as a motivation for our proposal.  
Namely, in the more common Riemannian case, when applied to the entire boundary system (thermal entropy), the HEE formula (\ref{EE}) for a bulk with a stationary AdS black hole gives us the same answer
as Wald's black hole entropy formula, because the extremal surface obtained by varying $S_{\text{HEE}}$ coincides with the horizon,
\cite{deBoer:2011wk}. A proof of the relation between Wald's black hole entropy formula and HEE in the Riemannian case can be found in \cite{Dong:2013qoa} and from there one can see that the relation can be more complicated in the case of general higher-curvature gravities.

Black hole entropy in CS gravity was briefly analyzed in \cite{Gallegos:2020otk}. One conclusion coming from this work is that the black hole entropy, including the contribution from axial torsion,
can be obtained by applying Wald's entropy formula, with a minor correction: we use a more general RC curvature instead of the usual Riemann curvature. In particular, for 5-dimensional CS gravity we have
\begin{align}
S_{\text{Wald}}
&=
2\pi
\int_{\mathrm{horizon}}
\left(
\frac{\partial L^{(5)}_{\text{CS}}}{\partial R^{ab}}
\right)
n^{ab}\\
     &=4\pi k\int_{\mathrm{horizon}} \varepsilon_{01ijk}n^{01}\left(R^{ij}e^k+\frac{1}{3 \ell^2}e^ie^je^k\right)\,,\nonumber
\end{align}
where $n^{01}$ is biorthogonal to the horizon of \eqref{metrika}, given by the same expression as in the Riemannian case. Note that this is the same type of substitution as in \eqref{anomm}.
Based on these observations, it seems reasonable to expect that a similar replacement $\mathcal{R}_{\gamma^{0}_{\text{ext}}}\to \mathcal{R}^{(\text{RC})}_{\gamma_{\text{ext}}}$ in (\ref{EE}) would lead to the same logarithmic torsional contribution as (\ref{ln1}). The agreement between the two results would make a strong support of the validity of our conjecture and motivate a starting point for a more general study of the entanglement entropy in a spacetime with torsion. Note, however, that in the proposed RC version of HEE, the variation of the entropy functional does not yield the same extremal surface as in the torsion-free case $(\gamma^{0}_{\text{ext}}\neq\gamma_{\text{ext}})$, and do not even 
necessarily yield the horizon as the extremal surface when applied to the whole boundary, thus differing from the RC version of Wald's entropy formula. We will further discuss this issue below.

Let us perform a concrete computation of the entanglement entropy of a solid cylinder in a QFT state dual to AdS$_{5}$ with axial torsion. Cylinder is interesting because its boundary's Riemannian curvature tensor vanishes, and thus no universal $\ln\epsilon$ term appears in \eqref{See4d}. The bulk line element takes the form of \eqref{metrika} for $\mu=0$, which can be written as 
\begin{equation}
    \diff s^2=\frac{\diff \rho^2}{4\rho^2}+\frac{1}{\rho}\left(-\diff t^2+\diff z^2+\diff r^2+r^2\diff\varphi^2  \right),
\end{equation}
where we assumed cylindrical parametrization $(z,r,\varphi)$ of the flat boundary. Note also that we work with static spacetime with no contortion components in the time direction, so that we can reduce our discussion to a $t$-constant slice. We take $\mathcal{A}$ to be a cylinder defined by $z\in[0,1]$, $\varphi\in[0,2\pi)$, $r\in[0,R]$, and consider a bulk spatial hypersurface defined by $r=f(\rho)$ with a unit normal 
\begin{equation}
n_\mu=\left(-\frac{f'(\rho)}{\sqrt{\rho+4\rho^2f'(\rho)^2}},0,\frac{1}{\sqrt{\rho+4\rho^2f'(\rho)^2}},0\right)\,.    
\end{equation}
The induced metric on the extremal surface is
\begin{equation}
d^{2}s_{\gamma_{\text{ext}}}=\left[\frac{1}{4\rho^2}+\frac{f'^{2}(\rho)}{\rho}\right]\diff \rho^2+\frac{1}{\rho}\left(\diff z^2+f^{2}(\rho)\diff\varphi^2  \right)\,,    
\end{equation}
with $\sqrt{h}=\frac{f(\rho)}{2\rho^{2}}\left[1+4\rho f'^{2}(\rho)\right]^{1/2}$.
It is now straightforward to calculate the induced RC curvature scalar by first focusing on a $t=\mathrm{const.}$ slice (which is trivially done in the case at hand) and then using the generalized RC version of the Gauss-Codazzi equation, 
\begin{align}
&\mathcal{R}^{(\text{RC})}_{\gamma_{\text{ext}}}=-\frac{2}{{f(\rho ) \left(4 \rho  f'(\rho )^2+1\right)^2}} \Big(48 \mathcal{C}^2 \rho ^3 f(\rho ) f'(\rho )^4\nonumber\\
    &+20 \rho  \left(\mathcal{C}^2 \rho +1\right) f(\rho ) f'(\rho )^2+\left(2 \mathcal{C}^2
   \rho +3\right) f(\rho )\\&+4 \rho ^2 f''(\rho )-16 \rho ^2 f'(\rho )^3+2 \rho  f'(\rho ) \left(8 \rho  f(\rho )
   f''(\rho )-1\right)\Big)\,.\label{Rrc}\nonumber
\end{align}
After evaluating action (\ref{EE}) we obtain
\begin{equation}\label{SeeCS}
S_{\text{HEE}}=4\pi^{2} k \int\frac{f(\rho ) \sqrt{4 \rho  f'(\rho )^2+1} }{2 \rho ^2}\left(1+\frac{1}{2}\mathcal{R}^{(\text{RC})}_{\gamma_{\text{ext}}}(\rho)\right)\textrm{d}\rho.
\end{equation} 
Now, in the purely Riemannian case, the profile $r=f(\rho)$ of the extremal surface can be obtained analytically as shown in \cite{deBoer:2011wk}, and its expansion in the neighborhood of the boundary $\rho=0$ is given by 
\begin{equation}\label{expansion}
r=R-\frac{\rho}{4R}+\mathcal{O}(\rho^{2})\,.    
\end{equation}
In the RC case, on the other hand, we cannot even determine the actual extremal surface with the assumed asymptotics. For that reason, we will try to use the torsion-free solution for the extremal surface. In addition, CS gravity is special. Namely, it turns out that in the torsion-free case we don't even need to know the coefficients of the $\rho^{n}$ ($n>0$) terms in the near-boundary expansion (\ref{expansion}) if we are only interested in the logarithmic term; we only need the boundary condition $r(\rho=0)=R$, which holds also in the RC case. This fact supports the Riemannian approximation for the extremal surface found in \cite{deBoer:2011wk}. 
It is hard to justify why this prescription would be correct in general. Nevertheless, given the result we obtained bellow, it seems plausible that the integration over the Riemannian surface even in the RC case is the right choice.  

By evaluating the RC HEE for 5-dimensional CS gravity (\ref{SeeCS}) on the Riemannian extremal surface (\ref{expansion}) we get 
\begin{equation}\label{konkretno}
    S_{\text{HEE}}=4\pi^{2} k\int \diff \rho\left(-\frac{R}{\rho^2}-\frac{\mathcal{C}^2R}{\rho}+\dots\,\right)\,,
\end{equation}
where we have omitted the finite terms. If we introduce a cut-off at $\rho=\epsilon^2$, we find a divergent contribution to the entanglement entropy of the form $    \mathcal{C}^2\ln\epsilon$,
in agreement with the expected result. The same conclusion holds if instead of a cylinder we take, for example, a ball, matching the result \eqref{ln1}.
\section{Discussion and outlook}

Ever since Einstein recognized gravity as a manifestation of spacetime curvature, it has been emphasized---most notably by Cartan---that there is no \emph{a priori} reason to restrict the affine connection to be torsion-free. In Einstein gravity, one assumes the Levi--Civita connection (torsion-free and metric-compatible), which is fixed uniquely by the metric. In RC geometry, by contrast, torsion is treated as an independent dynamical element of spacetime geometry. Although the experimental success of general relativity has rendered such extensions less central in phenomenology, interest in torsion has been steadily growing. In particular, torsion has emerged as a useful effective description in condensed-matter settings, \emph{e.g.}\ in models of spin transport \cite{Gallegos:2021bzp, Gallegos:2022jow, Erdmenger:2022nhz}, and it has also been argued to capture important aspects of holographic transport in 5-dimensional CS gravity \cite{Gallegos:2020otk}. These developments motivate extending holographic methods to RC backgrounds---including probes such as entanglement entropy---and treating torsion as a physically meaningful element of the theory.

In this paper, we proposed a method to compute the torsion-induced universal logarithmic term in the entanglement entropy of a 4-dimensional CFT holographically dual to an AdS$_5$ bulk with axial torsion, realized as a particular solution of 5-dimensional CS gravity. Our starting point was the known torsion-free (purely Riemannian) expression for the universal $\ln\epsilon$ contribution in a 4-dimensional CFT. Specifically, if the c-type central charge vanishes (which is the case for 4-dimensional CFT dual to 5-dimensional CS gravity), the entanglement entropy in the torsion-free sector is captured by the Ricci scalar induced on the entangling surface $\Sigma$ of a given spatial region $\mathcal{A}$ (\ref{See4d}).  
We then put forward the following prescription: promote the torsion-free Ricci scalar to its RC counterpart. This substitution lead to a universal logarithmic divergence of the form
\begin{equation}\label{Stor_discussion}
S_{\mathcal A}^{(\mathrm{torsion})}\ \sim\ A_{\Sigma}\,\mathcal{C}^{2}\,\ln\epsilon\,,
\end{equation}
where $\mathcal{C}$ is the strength of spatial torsion, so that the contribution vanishes in the torsion-free limit. 

Independently, motivated by the established relation between the holographic entanglement entropy and Wald's black-hole entropy formula in the torsion-free setting---and by the fact that Wald's construction admits a direct generalization to RC geometry---we proposed a second, \emph{a priori} distinct conjecture: at least for 5-dimensional CS gravity, the holographic entanglement entropy can be upgraded to incorporate torsion by the same replacement $\mathcal{R}\to\mathcal{R}^{(\mathrm{RC})}$. However, the resulting entropy functional does not admit an extremal surface with proper asymptotics, so we do the next best thing: evaluate the torsion-full entropy functional on the torsion-free extremal surface. This approximation can be further supported by the fact that, in 5-dimensional CS gravity, we only need the boundary condition to extract the logarithmic divergence. Remarkably, this second route reproduces precisely the same torsion-induced logarithmic contribution \eqref{Stor_discussion}. 
The agreement between these two computations therefore provides a nontrivial consistency check of our prescription and supports the conclusion that the RC Ricci scalar replacement correctly captures the universal logarithmic torsional contribution to entanglement entropy in this holographic model. It also suggests a plausible direction for extending holographic entanglement entropy beyond CS gravity.

We can further motivate the physical significance of the torsion-induced term \eqref{Stor_discussion} by a heuristic argument. Namely, similar universal logarithmic entanglement entropy terms are also found in even-dimensional massive QFT \cite{Lewkowycz:2012qr}; in $D=4$ we have   
$\sim A_{\Sigma}m^{2}\ln \epsilon$.
Now consider a 4-dimensional action, discussed in \cite{Aviles:2024muk}, for a real massless scalar field $\phi$ non-minimally coupled to a background described by $e^{a}$ and $\omega^{ab}$, 
\begin{align}
  \int \epsilon_{abcd}
\Bigg(
\phi^{2} R^{ab}
+ 
   \frac{1}{2} Z^{2}
e^{a} \wedge e^{b}
+ 4\phi Z^{a}T^{b}
\Bigg )\nonumber
\wedge e^{c} \wedge e^{d}\,.
\end{align}
where $Z^{a}=e_{\mu}^{a}\nabla^{\mu}\phi$. This action remains invariant under appropriate Weyl rescaling, induced from the bulk considerations in \cite{Cvetkovic:2017fxa}. 
Evaluating the action on our AdS$_5$ bulk with axial torsion, we get, up to an overall factor,
\begin{align}
  \sim\int\left[\eta^{\mu\nu}\partial_{\mu}\phi\partial_{\nu}\phi-\mathcal{C}^{2}\phi^{2}\right]\,,
\end{align}
which is the action for a massive scalar field, although with $m^{2}=-\mathcal{C}^{2}<0$. This might seem unsatisfactory at first sight. However, a more general definition of Weyl rescaling from \cite{Aviles:2024muk} would give the right sign of the mass term, though this Weyl rescaling would not be derived from the bulk considerations from \cite{Cvetkovic:2017fxa}. Of course, the boundary dual to CS gravity is more complicated than a non-interacting scalar theory, which simply exemplifies a Weyl invariant model sensitive to torsion that also has an intuitive interpretation of \eqref{Stor_discussion}: axial torsion can act as an emergent infrared scale in the boundary effective description, producing a universal logarithmic term in the entanglement entropy of the same form as in massive four-dimensional QFT, namely $\sim A_{\Sigma}\mathcal{C}^{2}\ln\epsilon$. 

To the best of our knowledge, the present work provides the first explicit holographic computation of the torsion-induced universal logarithmic contribution to entanglement entropy. While our construction relies on special features of CS gravity (in particular the relatively simple structure of HEE in the torsion-free case), it is natural to ask to what extent analogous torsion-induced universal terms persist in more general torsion-full holographic models. Clarifying this question---for example, by considering different classes of bulk actions or generalizing other known holographic methods to RC geometry---is an interesting direction for future work.
\vspace{-\baselineskip}

 \section*{Acknowledgements}
The work of D.D. and D.G. is supported by the funding provided by the Faculty of Physics, University of Belgrade, through grant number 451-03-47/2023-01/200162 by the Ministry of Science, Technological Development and Innovations of the Republic of Serbia. The research was supported by the Science Fund of the Republic of Serbia, grant number TF C1389-YF, Towards a Holographic Description of Noncommutative Spacetime: Insights from Chern-Simons Gravity, Black Holes and Quantum Information Theory - HINT. The authors thank the participants of HINT Open Day for a fruitful discussion on the topic presented in this paper.

\enlargethispage{3\baselineskip}

\bibliographystyle{elsarticle-num} 
\bibliography{ref}

\end{document}